\documentstyle[proceedings]{crckapb} 
\input{epsfig.sty}

\begin{opening}
\title{LEADING PARTICLES AND DIFFRACTIVE SPECTRA\protect\\
       IN THE INTERACTING GLUON MODEL}

\author{F.O.Dur\~aes}
\author{F.S.Navarra}
\institute{Instituto de Fisica, Universidade de S\~ao Paulo\\
C.P.66318, 05389-970 S\~ao Paulo, SP, Brazil}
\author{G.Wilk}
\institute{The Andrzej Soltan Institute for Nuclear Studies,\\
Nuclear Theory Deaprtment (Zd-PVIII)\\
ul. Ho\.za 69, 00-681 Warsaw, Poland}

\end{opening}

\runningtitle{Leading particles and diffractive spectra...}

\begin{document}


\begin{abstract}
We discuss the leading particle spectra and diffractive mass spectra
from the novel point of view, namely by treating them as particular
examples of the general energy flow phenomena taking place in the
multiparticle production processes. We argue that they show a high
degree of universality what allows for their simple description in
terms of the Interacting Gluon Model developed by us some time ago.
\end{abstract}

\section{Introduction}

The multiparticle production processes are the most complicated
phenomena as far as the number of finally involved degrees of freedom
is concerned. They are also comprising the bulk of all inelastic
collisions and therefore are very important - if not {\it per se}
than as a possible background to some other, more specialized
reactions measured at high energy collisions of different kinds. The
high number of degrees of freedom calls inevitably for some kind of
statistical descrition when addressing such processes. However, all
corresponding models have to be supplemented by information on the
fraction of the initial energy deposited in the initial object(s) 
(like fireball(s)) being then the subject of further investigations. 

Some time ago we have developed a model describing such energy
deposit (known sometimes as {\it inelasticity}) and connecting it
with the apparent dominance of multiparticle production processes by
the gluonic content of the impinging hadrons, hence its name: {\it
Interacting Gluon Model} (IGM) \cite{Fowler89}. Its classical application
to description of inelasticity \cite{Duraes93} and multiparticle production
processes in hydrodynamical model approach \cite{Duraes94} was soon
followed by more refined applications to the leading charm production
\cite{Duraes95} and to the (single) diffraction dissociation, both in
hadronic reactions \cite{Duraes97a} and in reactions originated by photons
\cite{Duraes97b}. These works allowed for providing the systematic
description of the leading particle spectra (which turned out to be
very sensitive to the presence of diffractive component in the
calculations, not accounted for before) \cite{Duraes98a} and clearly
demonstrated that they are very sensitive to the amount of gluonic
component in the diffracted hadron \cite{Duraes98b} and
\cite{Duraes98c}. We have found it 
amusing that all the results above were obtained using the same set
of basic parameters with differences arising essentially only because
of different kinematical limits present in each particular
application (i.e., in different allowed phase space). All this
points towards the kind of {\it universality} of energy flow patterns
in all the above mentioned reactions. 

Two recent developments prompted us to return again to the IGM ideas
of energy flow: one was connected with the new, more refined data
on the leading proton spectra  in $ep \rightarrow e'pX$ obtained
recently by ZEUS collaboration \footnote{Private information from
A.Garfagnini, see also ZEUS Collab. presentation in these
proceedings.} (which are apparently 
different from what has been used by us before in \cite{Duraes98a},
\cite{Duraes98b} and \cite{Duraes98c}).
The other was recent work on the central mass production in Double
Pomeron Exchange (DPE) process reported in \cite{Brandt02} allowing in
principle for deduction of the Pomeron-Pomeron total cross section
$\sigma_{I\! P-I\! P}$. In what follows we shall therefore provide a
brief description of IGM, stressing the universality of energy flow
it provides and illustrating it by some selected examples from our
previous works. The new results of ZEUS will be then shown again and
commented. Finally, we shall present our recent application of the
IGM to the DPE processes as well \cite{Duraes02}.

\section{IGM and some of its earlier applications}

The main idea of the model is that nucleon-nucleon collisions (or any
hadronic collisions in general) at high energies can be treated as an
incoherent sum of multiple gluon-gluon collisions, the valence quarks
playing a secondary role in particle production. While this idea is
well accepted for large momentum transfer between the colliding
partons, being on the basis of some models of minijet and jet
production (for example HIJING \cite{Wang92}), in the IGM  its
validity is extended down to low momentum transfers, only slightly
larger than $\Lambda_{QCD}$. At first sight this is not  justified
because at lower scales there are no independent gluons, but rather a
highly correlated configuration of color fields. There are, however,
some indications coming from lattice QCD calculations, that these
soft gluon modes are not so strongly correlated. One of them is the
result obtained in \cite{Giacomo92}, namely that the typical correlation
length of the soft gluon fields is close to $0.3$ fm. Since this
length is still much smaller than the typical hadron size, the gluon
fields can, in a first approximation, be treated as uncorrelated.
Another independent result concerns the determination of the typical
instanton size in the QCD vaccum, which turns out to be of the order
of $0.3$ fm \cite{Shaefer98}. As it is well known (and has been recently
applied to high energy nucleon-nucleon and  nucleus-nucleus
collisions) instantons are very important as mediators of  soft gluon
interactions \cite{Shuryak00}. The  small size of the active instantons
lead to short distance  interactions between soft gluons, which can
be treated as independent.

These two results taken together suggest that a collision between the
two gluon clouds (surrounding the valence quarks) may be viewed as a
sum of independent binary gluon-gluon collisions, which is the basic
idea of our model. The interaction follows then the usual IGM
picture \cite{Fowler89} and \cite{Duraes98a}, namely: the valence quarks fly through
essentially undisturbed whereas the gluonic clouds of both
projectiles interact strongly with each other (by gluonic clouds we
understand a sort of "effective gluons" which include also their
fluctuations seen as $\bar{q}q$ sea pairs) forming a kind of central
fireball (CF) of mass $M$. The two impinging projectiles (usually
protons/antiprotons and mesons) loose fractions $x$ and $y$ of their
original momenta and get excited forming what we call {\it leading
jets} (LJ's) carrying $x_p= 1 -x$ and $x_{\bar p}= 1 -y$ fractions of
the initial momenta. Depending on the type of the process under
consideration one encounters different situation depicted in Fig. 1.
In non-diffractive (ND) processes one is mainly interested only in CF
of mass $M$, in single diffractive (SD) ones in masses $M_X$ or $M_Y$
(comprising also the mass of CF) whereas in double Pomeron exchanges
(DPE) in a special kind of CF of mass $M_{XY}$. The only difference
between ND and SD or DPE processes is that in the later ones the
energy deposition is done by a restricted bunch of gluons which in
our language are forming what is regarded as a kind of "kinematical"
Pomeron ($I\!P$), the name which we shall use in what follows.

\begin{figure}[h]
\begin{center}
\epsfysize=5.5cm
\centerline{\epsfig{figure=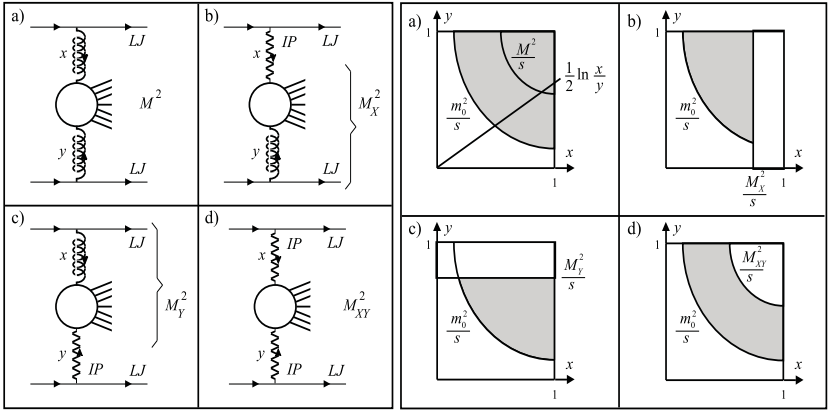,width=11cm}}
\caption{Left panel: schematic IGM pictures for $(a)$ non-diffractive
(ND), $(b)$ and $(c)$  single diffractive (SD) and $(d)$ double
Pomeron exchange (DPE) processes. Their corresponding phase space
limits are displayed on the right panel. The $\frac{1}{2}\ln
\frac{x}{y}$ line in the right $(a)$ panel indicates the rapidity $Y$
of the produced mass $M$.} 
\end{center}
\end{figure}

The central quantity in IGM is then the probability to form a CF 
carrying momentum fractions $x$ and $y$ of two colliding hadrons
\cite{Fowler89} and \cite{Duraes98a} which is given by:  
\begin{eqnarray}
\chi (x,y) &=&\frac{\chi _0}{2\pi \sqrt{D_{xy}}}\cdot
 \exp \left\{ -\frac 1{2D_{xy}}\,\left[ \langle y^2\rangle (x-\langle
x\rangle )^2\, +\, \langle x^2\rangle (y-\langle y\rangle )^2\right]
\right\} \nonumber \\ 
&\times& \exp \left\{-\frac 1{D_{xy}}\,\langle xy\rangle
(x-\langle x\rangle )(y-\langle y\rangle ) \right\} ,
\label{eq:CHI}
\end{eqnarray}
where $D_{xy} = \langle x^2\rangle \langle y^2\rangle -\langle
xy\rangle^2$ and
\begin{equation}
\langle x^n\, y^m\rangle =
\int_0^{x_{max}}\!dx'\,x'^n\,\int_0^{y_{max}}\!dy'\,y'^m\,\omega (x',y'),
\label{eq:defMOM}
\end{equation}
with $\chi _0$ defined by the normalization condition,
$\int_0^1\!dx\,\int_0^1\!dy\,\chi (x,y)\theta (xy-K_{min}^2)=1$, where
$K_{min}=\frac{m_0}{\sqrt{s}}$ is the minimal inelasticity defined 
by the mass $m_0$ of the lightest possible CF. The spectral function,
$\omega (x',y')$, contains all the dynamical input of the IGM. Their
soft and semihard components are given by (cf. \cite{Duraes93}):  
\begin{equation}
\omega(x',y') = \omega^{(S)}(x',y') + \omega^{(H)}(x',y')
\end{equation}
with
\begin{equation}
\omega^{(i)} (x',y')\,=\,\frac{\hat{\sigma}^{(i)}
_{gg}(x'y's)}{\sigma (s)}\,G(x')\,G(y')\,\theta \left(
x'y'-\left[K^{(i)}\right]_{min}^2\right) ,  \label{eq:OMEGAS} 
\end{equation}
where $i=S,H$, and $K^{(S)}_{min} = K_{min}$ whereas $K^{(H)}_{min} =
2p_{T_{min}}/\sqrt{s}$. The values of $x_{max}$ and $y_{max}$ depend
on the type of the process under consideration (cf. Fig. 2). For ND
processes $x_{max}=y_{max}=1$ (all phase space above the minimal one
is allowed) whereas for SD and DPE there are limitations seen in Fig.
2 and discussed in more detail in the appropriate sections below.
Here $G$'s denote the effective number of gluons from the
corresponding projectiles (approximated in all our works by the
respective gluonic structure functions), $\hat{\sigma}^S _{gg}$ and
$\hat{\sigma}^H _{gg}$ are the soft and semihard gluonic cross
sections, $p_{T_{min}}$ is the minimum transverse momentum for
minijet production and $\sigma$ denotes the impinging projectiles
cross section. 

Lets us close this section by mentioning that, as has been shown in
\cite{Fowler89}, \cite{Duraes94} and \cite{Duraes95}, IGM can describe both the hadronic and
nuclear collision data (providing initial conditions for the Landau
Hydrodynamical Model of hadronization \cite{Duraes94}) as well as some
peculiar, apparently unexpected features in the leading charm
production \cite{Duraes95} (mainly via its strict energy-conservation
introducing strong correlations between production from the CF and
LJ). It was done with the same form of the gluonic structure
function in the nucleon used: $G(x) = p(m+1)(1-x)^m/x$ with defold
value of $m=5$ and with the fraction of the energy-momentum allocated
to gluons equal to $p=0.5$ and with $\sigma^{(i)}_{gg}(xys)=const$
(notice that results are sensitive only to the combination of
$p^2\sigma_{gg}/\sigma$). 

\section{Single Diffractive processes in the IGM}

In the last years, diffractice scattering processes received
increasing attention mainly because of their potential ability to
provide information about the most important object in the Regge
theory, namely the Pomeron ($I\!P$), its quark-gluonic structure and
cross sections. Not entering the whole discussion \cite{Goulianos83}
\footnote{See also talk by K.Goulianos in these proceedings.} we would
like to show here the possible approach, based on the IGM, towards
the mass($M_X$) distributions provided by different experiments. In
Figs. 1b and 1c the understanding of what such mass means in terms if
the IGM is clearly shown: it contains both the central fireball and
the LJ from the initial  particle which got excited. The only
difference between it and, say, the corresponding object which could
be formed also in Fig. 1a is that the energy transfer from the
diffracted projectile is now done by the highly correlated bunch of
gluons dennoted $I\!P$ which are supposed to be in the colour singlet
state. The other feature also seen in Fig. 1 (right panel) is that
now only the limited part of the phase space supporting the
$\chi(x,y)$ distribution is allowed and that the limits depend on the
mass $M_X$ we are going to produce (and observe). 

\begin{figure}[h]
\begin{center}
\epsfysize=5cm
\centerline{\epsfig{figure=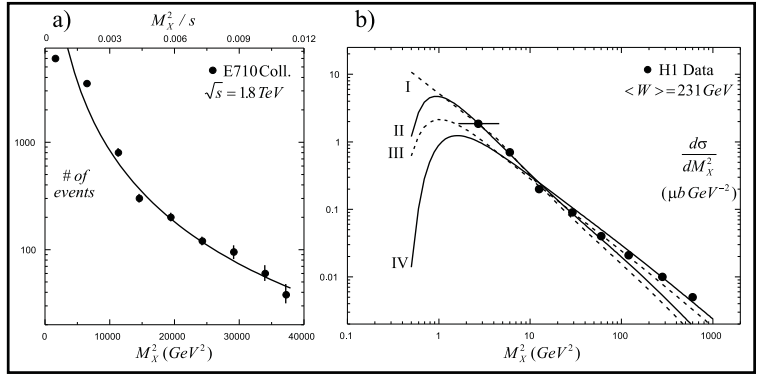,width=10cm}} 
\caption{Examples of diffractive mass spectra for $(a)$ $p\bar{p}$
collisions compared with Tevatron data \protect\cite{Abe94} (Fig. 4 from
\protect\cite{Duraes97a}) and for $(b)$ $\gamma p$ collisions compared with H1 data
\protect\cite{Adloff97} (Fig. 2b from \protect\cite{Duraes97b}, Vector Dominance Model has been
used here with $G^{\rho}(x) = G^{\pi}(x) = 2p(1-x)/x$; the
different curves correspond to different choices of $(m_0$
GeV$,\sigma$ mb $)$: I$=(0.31,2.7)$, II$=(0.35,2.7)$,
III$=(0.31,5.4)$ and IV$=(0.35,5.4)$). } 
\end{center}
\end{figure}

In technical terms it means that in comparison to the previous
applications of the IGM we are free to change both the possible shape
of the function $G_{I\!P}(x)$ (telling us the number of gluons
participating in the process) and the cross section $\sigma$ in the
spectral function $\omega$ in eq. (\ref{eq:OMEGAS}) above. Actually
we have found that we can keep the shape of $G(x)$ the same as before
and the only change necessary (and sufficient) to reach agreement
with data is the amount of energy-momentum $p=p_{I\!P}$ allocated to
the impinging hadron and which will find its way in the object we
call $I\!P$. It turns out that $p_{I\!P} \simeq 0.05$ (to be compared
with $p\simeq 0.5$ for all gluons encountered so far). In Fig. 3 we
provide a sample of results taken from \cite{Duraes97a} and
\cite{Duraes97b}. They all have 
been obtained by putting $x_{max}=1$ and $y_{max}=M_X^2/s$ in eq.
(\ref{eq:defMOM}) above and by writing
\begin{equation}
\frac{dN}{dM_X^2}\, =\, \int_0^1\!\!dx\int_0^1\!\! dy\,
\chi(x,y)\delta\left(M^2_X-sy\right)\Theta\left(xy - K_{min}^2\right).
\label{eq:SDD}
\end{equation}
As can be explicitely shown the characteristic $1/M^2_X$ behaviour of
diffractive mass spectra are due to the $G(x) \sim 1/x$ behaviour of
the gluonic structure functions for small $x$. The full formula
results in small deviations following precisely the trend provided by
experimental data (and usually attributed in the Regge model approach
to the presence of additional Reggeons \cite{Goulianos83}).

\section{Leading Particle spectra in the IGM}

With the above development of the IGM one can now think about the
systematic survey of the leading particle spectra, both in hadronic
and in $\gamma p$ collisions \cite{Duraes98a} (cf. Figs. 4 and 5). The
specific prediction of the IGM connected with the amount of gluons in
the hadron available for interactions (i.e., for the slowing down of 
the original quark content of the projectile) has been discussed in
\cite{Duraes97a}, \cite{Duraes97b} and \cite{Duraes98b},
\cite{Duraes98c}, and shown here in Fig. 6. The leading particle 
can emerge from different regions of the phase space (cf. Figs. 1)
and distribution of its momentum fraction $x_L$ is given by \cite{Duraes97a}: 
\begin{eqnarray}
F(x_L)\, =\, &&(1-\alpha)\int^1_{x_{min}}\!\! dx\,
\chi^{(nd)}\left(x;y=1-x_L\right)\, + \nonumber\\
&& +\, \sum_{j=1,2}\alpha_j
\int^1_{x_{min}}\!\! dx\, \chi^{(d)}\left(x;y=1-x_L\right) ,\label{eq:FLP}
\end{eqnarray}
\begin{figure}[h]
\begin{center}
\epsfysize=10cm
\centerline{\epsfig{figure=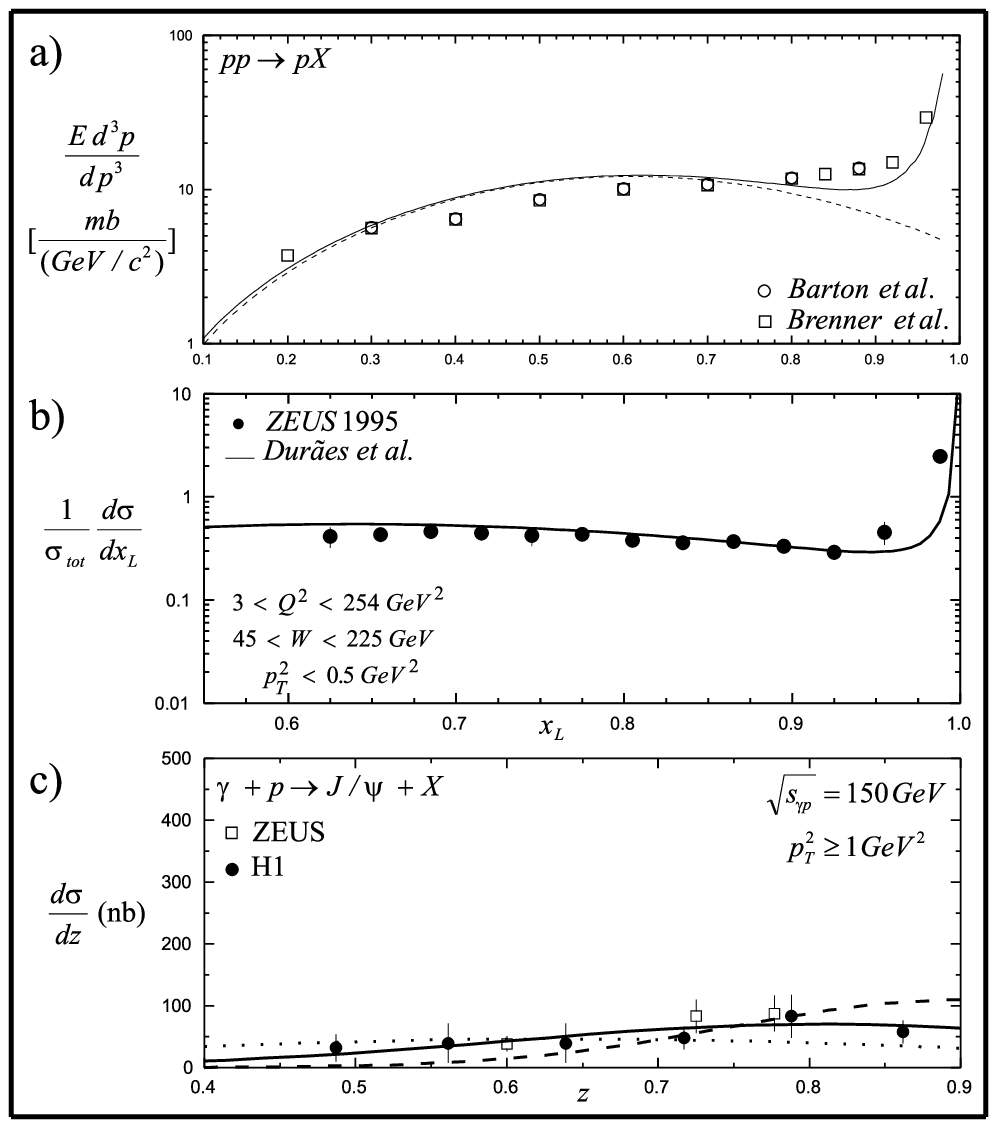,width=10cm}}
\caption{
$(a)$ Example of comparison of our LP spectra $F(x_L)$ with
data  from \protect\cite{Barton83} and \protect\cite{Brenner82} (Fig.
2a from \protect\cite{Duraes97a}). 
$(b)$ Comparison between our calculation and and the new data on the
leading proton spectrum measured at HERA by the ZEUS Collab. 
$(c)$ Fits to leading $J/\Psi$ spectra as given by ZEUS and H1 groups
\protect\cite{Aid96}, \protect\cite{Derrick95} and
\protect\cite{Breitweg97} (cf. \protect\cite{Duraes98b} and 
\protect\cite{Duraes98c}; the fixed value of 
$\sigma^{inel}_{J/\Psi-p} = 9$ mb has been used and results for three
different choices of $p^{J/\Psi}$ are displayed: $0.066$ - dashed
line, $0.033$ - solid line and $0.016$ -dotted line).
}
\end{center}
\end{figure}
where $\alpha = \alpha_1 + \alpha_2$ is the total fraction of single
diffractive  $(d)$ events (from the upper and lower legs in Fig. 1,
respectively, both double DD and DPE events are neglected here) and where
\begin{equation}
x_{min} = {\rm Max}\left[\frac{m_0^2}{(1-x_L)s};
\frac{\left(M_{LP}+m_0)\right)^2}{s}\right]  \label{eq:xmin}
\end{equation}
with $M_{LP}$ being the mass of the LP under consideration. Notice
that the $\alpha$ is essentially a new parameter here, which should
be of the order of the ratio between the total diffractive and total
inelastic cross sections \cite{Duraes97a}. All other parameters leading to
results in Fig. 4 are the same as established before \footnote{In
what concerns comparison with ZEUS data see also presentation of ZEUS
Collab. in these proceedings. Our present results differ from Fig. 4
in \cite{Duraes97a} where the preliminary ZEUS data were used
instead. The only difference between the two fits is that whereas in
\protect\cite{Duraes97a} we were assuming that $30\%$ of the LP comes
from diffraction, now it is only $10\%$.}. 

We want to stress here the fact that the fair agreement with data
observed in the examples  shown in Figs. 4 and 5 is possible only
because the diffraction processes have been properly incorporated in
calculating the LP spectra \cite{Duraes97a}. As far as the energy flow is
concerned the IGM works extremely well (including the pionic LP not
shown here but discussed in \cite{Duraes97a}) with essentially two
parameters only: the nonperturbative gluon-gluon cross section and the
fraction of diffractive events. At the same time, assuming the Vector
Dominance Model and replacing impinging photon by its hadronic
component (as in Fig. 3b), we are able to describe also the leading
proton spectra observed in $e-p$ reactions. Also here the inclusion
of diffractive component turns out to be crucial factor to get
agreement with data. We can also describe fairly well pionic LP (not
shown here, cf. \cite{Duraes97a}) and the observed differences turns out to
be due to their different gluonic distributions. The crucial role
played by the parameter $p$ representing the energy-momentum fraction
of a given hadron allocated to gluons is best seen in the Fig. 3c
example showing fit to data for leading $J/\Psi$ photoproduction
\cite{Aid96}, \cite{Derrick95} and \cite{Breitweg97}. It turns out
that (after accounting for the proper kinematics in (\ref{eq:xmin})
and presence of diffraction processes 
as discussed above) the only parameter to which results are really
sensitive is $p=p^{J/\Psi}$ which, as shown in Fig. 3c, has to be
astonishingly small, $p^{J/\Psi} = 0.033$. However, closer scrutiny
shows us that this is exactly what could be expected from the fact
that charmonium is a non-relativistic system and almost all its mass
comes from the quark masses leaving therefore only small fraction,
\begin{equation}
p^{J/\Psi}\, =\, \frac{M_{J/\Psi} - 2 m_c}{M_{J/\Psi}}\, \simeq \,
0.033 ,\label{eq:JPSIi}
\end{equation}
for gluons (here $m_c = 1.5$ GeV and $M_{J/\Psi} = 3.1$ GeV).

\section{Double Pomeron Exchange in the IGM}

Our latest application of the IGM discussed here will be for the DPE
processes seen as a specific energy flow (cf. Fig. 1 d) taking place
from both colliding particles and directed into the central region.
The difference between it and the "normal" energy flow as represented
by Fig. 1a is that now the gluons involved in this process must be
\begin{figure}[h]
\begin{center}
\epsfysize=5cm
\centerline{\epsfig{figure=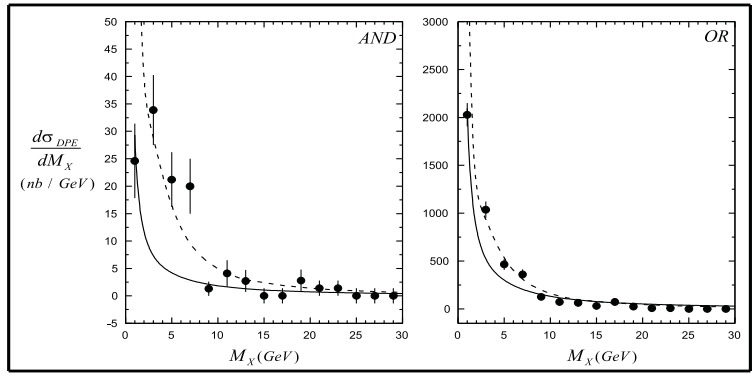,width=10cm}}
\caption{Our fits to two types of DPE diffractive mass distribution
given by \protect\cite{Brandt02} with $\sigma_{I\!P-I\!P} = 1$ mb (solid
lines) and $0.5$ mb (dashed lines).}
\end{center}
\end{figure}
confined to what is usually refered to as Pomeron ($I\!P$). Such
process war recently measured by UA8 \cite{Brandt02} and used (using the
normal Reggeon calculus arguments) to deduce the $I\!P\!-\!I\!P$ cross
section, $\sigma_{I\!P\!-\!I\!P}$. It turned out that using this
method one gets $\sigma_{I\!P\!-\!I\!P}$ which apparently depends on
the produced mass$M_{XY}$. This fact was tentatively interpreted as
signal of glueball formation \cite{Brandt02}.  However, when seen from the
IGM point of view mentioned above, where 
\begin{equation}
\frac{1}{\sigma}\frac{d\sigma}{dM_{XY}}\, =\,
\frac{2M_{XY}}{s}\int^1_{\frac{M^2_{XY}}{s}}\frac{dx}{x}\,
\chi\left(x,y=\frac{M^2_{XY}}{xs}\right)\, \Theta\left(M^2_{XY} -
m_0^2\right) , \label{eq:DPE} 
\end{equation}
the \cite{Brandt02} data can be fitted (see Fig. 4) with the same set
of parameters as used previously to describe the SD processes
\cite{Duraes97a} and \cite{Duraes97b} and with constant value of
$\sigma_{I\!P-I\!P}=0.5$ mb (which is new parameter here). No
glueball concept is needed here. 

\section{Summary and conclusions}

The picture which is emerging from the above discussion is that the
energy flows, which are present in all multiparticle production
reactions, are apparently a kind of universal phenomenon in the
following sense: they are all sensitive mainly to the gluonic content
of the colliding projectiles (i.e., both to the number of gluons
as given by the form of the function $G(x)$ and to the amount of
energy-momentum of the hadron, $p$, they carry and to the gluonic
cross section which defines the actual effectiveness of the gluonic
component in the energy transfer phenomenon). Their sensitivity to 
other aspects of the production process (except of kinematical limits
provided by the observed masses, as illustrated in Fig. 1) is only of
secondary importance.

To close our arguments two other applications of IGM should be,
however, tested: whether it can also describe the final, yet
unchecked energy flow pattern as the one provided by the Double
Diffraction Dissociation processes and whether it can be applied in
such simple form also to reactions with nuclei (first attempts were
already done at the very beginning of the history of IGM in
\cite{Fowler89}, but they were too crude to be convincing at present). We
plan to address these questions elsewhere.\\

\underline{Acknowledgements}: This work has been supported by FAPESP,
CNPQ (Brazil) (FOD and FSN) and by KBN (Poland) (GW). One of us (GW)
would like to thank also the Bogolubow-Infeld Program (JINR, Dubna)
for financial help in attending the Diffraction 2002 conference where
the above material has been presented.


\begin{thebibliography}{}


\bibitem[\protect\citeauthoryear{Abe {\it et al.}}{1994}]{Abe94}
Abe, F. et al. (CDF Collab.) (1994) Measurement of $p\bar{p}$ single
difraction dissociation at $\sqrt{s}=546$ and $1800$ GeV,
{\it Physical Review D}, {\bf Vol. ~no. 50}, pp.~ 5535-5590
              
\bibitem[\protect\citeauthoryear{Adloff {\it et al.}}{1997}]{Adloff97}
Adloff, C. et al. (1997) Difraction dissociation in photoproduction
at HERA, {\it Zeitschrift f\"ur Physik C}, {\bf Vo. ~no.~74}, 221-235

\bibitem[\protect\citeauthoryear{Aid {\it et al.}}{1996}]{Aid96}
Aid, S. et al. (H1 Collab.) (1996) Elastic and inelastic
photoproduction of $J/\Psi$ mesons at HERA, {\it Nuclear Physics B},
{\bf Vo. ~no.~472}, 3-31

\bibitem[\protect\citeauthoryear{Barton {\it et al.}}{1983}]{Barton83}
Barton, D.S. et al. (1983) Experimental studies of the $A$ dependence
of inclusive hadron fragmentation, {\it Physical Review D}, {\bf Vol.
~no. ~27}, 2580-2599

\bibitem[\protect\citeauthoryear{Brandt {\it et al.}}{2002}]{Brandt02}
Brandt A., {\it et al.} (2002) A Study of Inclusive
Double-Pomeron-Exchange in $p\bar{p} \rightarrow pX\bar{p}$ at
$\sqrt{s} = 630\,GeV$, {\it The European Pysical Journal C}, {\bf
Vol.~no. 25}, pp.~ 361-377

\bibitem[\protect\citeauthoryear{Breitweg {\it et al.}}{1997}]{Breitweg97}
Breitweg, J. et al. (ZEUS Collab.) (1997) Measurements of inelastic
$J/\Psi$ photoproduction at HERA, {\it Zeitschrift f\"ur Physik C},
{\bf Vo. ~no.~76}, 599-612

\bibitem[\protect\citeauthoryear{Brenner {\it et al.}}{2002}]{Brenner82}
Brenner, A.E. et al. (1982) Experimental study of single-particle
inclusive hadron scattering and associated multiplicities, {\it
Physical Revie D}, {\bf Vol. ~no. ~26}, 1497-1553

\bibitem[\protect\citeauthoryear{Derrick {\it et al.}}{1995}]{Derrick95}
Derrick, M. et al. (ZEUS Collab.) (1995) Measurement of the cross
section for the reaction $\gamma p \rightarrow J/\Psi p$ with the
ZEUS Detector at HERA, {\it Physics Letters B}, {\bf Vo. ~no. ~350},
120-134 

\bibitem[\protect\citeauthoryear{Duraes {\it et al.}}{1993}]{Duraes93}
Dur\~aes, F.O., Navarra, F.S. and Wilk, G. (1993) Minijets and
the behavior of inelasticity at high energies, {\it Physical Review
D}, {\bf Vol. ~no.~47}, pp.~3049-3052

\bibitem[\protect\citeauthoryear{Duraes {\it et al.}}{1994}]{Duraes94}
Dur\~aes, F.O., Navarra, F.S. and Wilk, G. (1994)
Hadronization and inelasticities, {\it Physical Review
D}, {\bf Vol.~no.~ 50}, pp.~6804-6810

\bibitem[\protect\citeauthoryear{Duraes {\it et al.}}{1995}]{Duraes95}
Dur\~aes, F.O., Navarra, F.S. and Wilk, G. (1995)
Hadronization and inelasticities, {\it Physical Review
D}, {\bf Vol.~no.~ 53}, pp.~6136-6143

\bibitem[\protect\citeauthoryear{Duraes {\it et al.}}{1997a}]{Duraes97a}
Dur\~aes, F.O., Navarra, F.S. and Wilk, G. (1997)
Diffractive dissociation in the interacting gluon model, {\it
Physical Review D}, {\bf Vol.~no.~ 55}, pp.~2708-2717

\bibitem[\protect\citeauthoryear{Duraes {\it et al.}}{1997b}]{Duraes97b}
Dur\~aes, F.O., Navarra, F.S. and Wilk, G. (1997)
Diffractive mass spectra at DESY HERA in the interacting gluon model,
{\it Physical Review D}, {\bf Vol.~no.~ 56}, pp.~R2499-R2503

\bibitem[\protect\citeauthoryear{Duraes {\it et al.}}{1998a}]{Duraes98a}
Dur\~aes, F.O., Navarra, F.S. and Wilk, G. (1998)
Systematics of leading particle production, {\it Physical Review
D}, {\bf Vol.~no.~ 58}, pp.~094034-1~-~094034-12

\bibitem[\protect\citeauthoryear{Duraes {\it et al.}}{1998b}]{Duraes98b}
Dur\~aes, F.O., Navarra, F.S. and Wilk, G. (1998)
$J/\Psi$ elasticity distribution in the vector dominance approach,
{\it Modern Physics Letters A}, {\bf Vol.~no.~ 13}, pp.~2873-2885

\bibitem[\protect\citeauthoryear{Duraes {\it et al.}}{1998a}]{Duraes98c}
Dur\~aes, F.O., Navarra, F.S. and Wilk, G. (1998)
Leading particle effect in the $J/\Psi$ elasticity distribution, {\it
Nrazilian Journal of Physics}, {\bf Vol.~no.~ 28}, pp.~505-509

\bibitem[\protect\citeauthoryear{Duraes {\it et al.}}{2002}]{Duraes02}
Dur\~aes, F.O., Navarra, F.S., G. Wilk, G. (2002) Extracting the
Pomeron-Pomeron Cross Section from  Diffractive Mass Spectra, 
ArXiv:hep-ph/0209140

\bibitem[\protect\citeauthoryear{Fowler {\it et al.}}{1989}]{Fowler89}
Fowler, G.N., Navarra, F.S., Pl\"umer, M., Vourdas, A., Weiner,
R.M. and Wilk, G. (1989) Interacting gluon model for hadron-nucleus
and nucleus-nucleus collisions in the central rapidity region, {\it
Physical Review C}, {\bf Vol. ~no.~40}, pp.~1219-1233  

\bibitem[\protect\citeauthoryear{Giacomo {\it et al.}}{1992}]{Giacomo92}
di Giacomo, A. and Panagopoulos, H. (1992) Field strength
correlations in the QCD vacum,  {\it Physics Letters B}, {\bf Vol.
~no.~285}, pp. ~133136

\bibitem[\protect\citeauthoryear{Goulianos}{1983}]{Goulianos83}
Goulianos, K. (1983) Diffractive interactions of hadrons at high
energies, {\it Physics Reports}, {\bf Vol. no.~~101}, pp. ~169-219

\bibitem[\protect\citeauthoryear{Shaefer {\it et al.}}{1998}]{Shaefer98}
Schaefer, T. and  Shuryak, E.  (1998) Instantons in QCD, {\it  Review
of Modern Physics}, {\bf Vo.~no.~70}, pp.~323-425

\bibitem[\protect\citeauthoryear{Shaefer}{1998}]{Shuryak00}
Shuryak, E.V. (2000)Towards the non-perturbative description of high
energy processes, {\it Physics Letters B}, {\bf Vol. no.~486}, pp.~
378-384 

\bibitem[\protect\citeauthoryear{Wang {\it et al.}}{92}]{Wang92}
Wang, X.N. and Gyulassy, M. (1992) Systematic study of particle
production in $p+(\bar{p})$ collisions via the HIJING model,
{\it Physical Review D} {\bf Vol. no.~45}, pp.~ 844-856

\end{thebibliography}
\end{document}